\documentclass[a4paper, 10pt, conference]{ieeeconf}      

\IEEEoverridecommandlockouts                             

\usepackage[pdftex]{graphicx}
\usepackage{amsmath,amssymb,enumerate,bbm,comment}
\usepackage[linesnumbered,vlined,ruled,commentsnumbered]{algorithm2e}
\usepackage{array}
\usepackage{xcolor}
\usepackage{aliascnt,hyperref,cleveref}

\def\R{\mathsf{R}}

\def\nset{\mathbb{N}}
\SetKwInput{KwInput}{Input}
\SetKwInput{KwOutput}{Output}

\def\eqdef{:=}
\def\eqsp{\;}

\def\1{\mathbbm{1}}
\def\Prox{\mathrm{Prox}}
\def\Z{\mathsf{Z}}

\def\R{\mathsf{R}}

\def\O{\mathsf{O}}

 \def\param{ \mathsf{R}}
\def\pas{\gamma}
\def\D2{\mathsf{D}_2}
\def\barD2{\overline{\mathsf{D}}_2}

\def\I{\mathsf{Id}}

\graphicspath{
{Figures/}
}

\title{\LARGE \bf
Temporal evolution of the Covid19 pandemic reproduction number: Estimations from proximal optimization to Monte Carlo sampling
}

\author{Patrice Abry$^{1}$, Gersende Fort$^{2}$, Barbara Pascal$^{3}$, Nelly Pustelnik$^{1}$
\thanks{$^{1}$CNRS, ENS de Lyon, Laboratoire de Physique, Lyon, France.
        {\tt\small firstname.lastname@ens-lyon.fr}}%
\thanks{$^{2}$CNRS, Institut de Math\'ematiques de Toulouse, Toulouse, France.
        {\tt\small gersende.fort@math.univ-toulouse.fr}. Part of this work is supported by the {\it Fondation Simone et Cino Del Duca, Institut de France.}}%
        \thanks{$^{3}$Univ. Lille, CNRS, Centrale Lille, CRIStAL, Lille, France.
        {\tt\small barbara.pascal@univ-lille.fr}}%
}

\begin{document}
\maketitle
\thispagestyle{empty}
\pagestyle{empty}

\begin{abstract}
Monitoring the evolution of the Covid19 pandemic constitutes a
critical step in sanitary policy design.  Yet, the assessment of the
pandemic intensity within the pandemic period remains a challenging
task because of the limited quality of data made available by public
health authorities (missing data, outliers and pseudoseasonalities,
notably), that calls for cumbersome and ad-hoc preprocessing
(denoising) prior to estimation.  Recently, the estimation of the
reproduction number, a measure of the pandemic intensity, was
formulated as an inverse problem, combining data-model fidelity and
space-time regularity constraints, solved by nonsmooth convex proximal
minimizations.  Though promising, that formulation lacks robustness
against the limited quality of the Covid19 data and confidence
assessment.  The present work aims to address both limitations: First,
it discusses solutions to produce a robust assessment of the pandemic
intensity by accounting for the low quality of the data directly
within the inverse problem formulation.  Second, exploiting a Bayesian
interpretation of the inverse problem formulation, it devises a Monte
Carlo sampling strategy, tailored to a nonsmooth log-concave \textit{a
  posteriori} distribution, to produce relevant credibility
interval-based estimates for the Covid19 reproduction number.
\newline
\indent \textit{Clinical relevance}
Applied to daily counts of new infections made publicly available by the Health Authorities for  around 200 countries, the proposed procedures permit robust assessments of the time evolution of the Covid19 pandemic intensity, updated automatically and on a daily basis. 
\end{abstract}

\section{Introduction}
\label{sec-intro}

\noindent {\bf Context.} \quad The online and daily surveillance of
the Covid19 pandemic intensity has become a critical societal stake
and constitutes a key preliminary step in the implementation of any
counter-measures by public authorities.  The evolution of the pandemic
is usually assessed from epidemiological models fed by daily counts of
new infections or death cases, the core data of any pandemic
surveillance strategy.  At the outbreak of the Covid19 pandemic,
facing the urgent need for data to monitor its evolution, significant
efforts were devoted by most national public health authorities to
collect such data and to make them publicly available.  However,
because of the emergency and sanitary crisis contexts, the available
data were of low quality, strongly corrupted by missing samples,
outliers and pseudo-seasonalities.  More surprisingly, after two years
of pandemic, the data collected by most countries remain of very
limited quality.  That low quality of the available data combined with
the need for online and regular (ideally daily) monitoring turn the
assessment of the pandemic intensity evolution into a far more
difficult task than when performed once the pandemic is over and with
consolidated data.  Further, assessing the confidence that can be
granted to such estimates also provides another critical and difficult
challenge.  The within pandemic online and daily updated estimation,
via credibility intervals, of the pandemic intensity from limited
quality data thus constitutes the core issue of this work. \\
\noindent {\bf Related work.} \quad Pandemic surveillance can be achieved with a large variety of tools, different in nature \cite{arino2021describing}. 
Estimating retrospectively the pandemic evolution, when it is over and after the data have been post-processed, is usually performed with {\it compartmental models} \cite{Liu2018,BCF}.
They suffer yet from heavy computational costs and are of limited robustness against the low quality of the Covid19 data. 
Instead, the pandemic intensity can be measured by the reproduction number, $R$,
that quantifies the number of second infections stemming from one same primary infection (cf. e.g.,
\cite{Diekmann1990,vandenDriessche2002,obadia2012,cori2013new}).
It has recently been proposed that relevant estimates of the time
evolution of the reproduction number can be obtained from nonsmooth
convex optimization procedures \cite{abry2020spatial}, with the
functional to minimize built from a pandemic model \cite{cori2013new}.
Attempts to create credibility intervals from a Bayesian
interpretation of that model complemented with Monte Carlo sampling
schemes were recently reported in \cite{artigas2021credibility} (see also \cite{abry:etal:2022}).
Though delivering epidemiologically realistic assessments of the
temporal evolution of the pandemic intensity, there is still a
significant need to increase the robustness of these tools against the
limited quality of the Covid19 data. \\
\noindent {\bf Goals, contributions and outline.} \quad Elaborating on \cite{artigas2021credibility,Pascal2021}, the goal of the present work is to further improve estimation robustness against the limited quality of the Covid19 data.
To that end, the reproduction number-based epidemiology model \cite{cori2013new} is recalled in Section~\ref{sec-Rmodel}. 
Section~\ref{sec-Resta} recalls the model-based and regularized inverse problem formulations for the estimation of $R$.
Sections~\ref{sec-Restb} and \ref{sec-Restc}  detail how the regularized inverse problem formulations can be modified to bring robustness against the  Covid19 data limited quality, while remaining close to the original epidemiological model. 
Section~\ref{sec-Restd} details the proposition of a construction of credibility interval-based estimation of $R$ relying on an original Markov Chain Monte Carlo sampler, referred to as \emph{Metropolis Adjusted Proximal-Gradient Algorithm}, refining  classical Metropolis Adjusted Langevin procedures. 
Using real Covid19 data described in Section~\ref{sec-data}, 
Section~\ref{sec-Res} discusses the performance of the proposed estimation procedures, for different countries. 

\section{Reproduction number model and estimation}
\label{sec-R}

\subsection{Pandemic model}  
\label{sec-Rmodel}
The pandemic model developed in \cite{cori2013new} and used here,
assumes that the count of daily new infections at time $t$, $Z_t$, is
drawn from a Poisson distribution, conditionally to past counts
$\Z_{1:t-1} \eqdef \{ Z_1,\ldots,Z_{t-1} \}$.  It further
postulates that the Poisson parameter, $p_t$, varies along time, and
depends on past counts $\Z_{1:t-1}$, on the causal \emph{serial
  interval} function $ \Phi_t$ and on the reproduction number at time
$t$, $R_t$: $ p^{(0)}_t \eqdef R_t \times \sum_{s = 1}^{\tau_\phi}
\Phi_s Z_{t-s}$.
The function $\Phi \eqdef (\Phi_t)_{t \geq 1}$ models the main epidemic evolution mechanism: the random delays between the onsets of symptoms in a primary and secondary cases 
\cite{cori2013new,obadia2012,thompson2019,Liu2018}.  For the Covid19
pandemic and for earlier pandemics of same types, it was shown that
$\Phi$ can be approximated as a Gamma function, with shape and rate
parameters corresponding to mean and standard deviation of 6.6 and 3.5
days, indicating a high risk of infecting other persons from 3 to 10
days after the symptoms have appeared
\cite{shujuan,Riccardo2020,Guzzetta}.

\subsection{Model-based estimation}  
\label{sec-Resta}

\noindent {\bf Maximum Likelihood estimation.} A natural estimation
strategy is based on maximizing the log-likelihood of the data.
Because of the Poisson distribution assumption in the model above, the
negative log-likelihood (also referred to as the data fidelity term)
is essentially a sum in time of the standard Kullback-Leibler
divergence, ${\cal L}(\param \lvert p^{(0)}) \eqdef \sum_{t = 1}^T
d_{\mathtt{KL}}(Z_t \lvert p^{(0)}_t)$ with $ d_{\mathtt{KL}}(z \lvert
p) \eqdef z \ln \frac{z}{p} + p - z $ when $ z > 0, p > 0$, $
d_{\mathtt{KL}}(z \lvert p) \eqdef p $ when $ z = 0, p \geq 0$ and $
d_{\mathtt{KL}}(z \lvert p) \eqdef +\infty$ otherwise, leading to:
\begin{align}
\label{eq:estRMLE}
\widehat \param^{(0)} \eqdef \underset{\param}{\mathrm{argmin}} \, {\cal L}(\param \lvert  p^{(0)}). 
\end{align}
Explicit calculations yield a simple closed-form expression: $\widehat R^{(0)} _t \eqdef Z_t /\sum_{s = 1}^{\tau_\phi} \Phi_s Z_{t-s}$. 
Both because of the low quality of the data, and of its being ill-conditioned (one new observation $Z_t$ is available to estimate daily $R_t$), $\widehat \param^{(0)}$ turns out to be extremely irregular along time (cf. Fig.~\ref{fig-Rfull}, second rows) and thus useless for epidemic surveillance. 

\noindent {\bf Penalized Maximum Likelihood estimation.} 
To favor temporal regularity in the estimate of $\param$, it was
proposed in \cite{abry2020spatial} to complement the data fidelity
term ${\cal L}(\param \lvert p^{(0)} )$ with a regularization term
based on the $L^1$-norm of the Laplacian of $\param$, $ \lVert
\boldsymbol{\mathrm{D}}_2 \param \rVert_1 \eqdef \sum_{t=3}^{T}
|R_{t-2}-2R_{t-1}+R_{t}|$:
\begin{align}
\label{eq:estRT}
\widehat \param^{(1)} \eqdef \underset{\param}{\mathrm{argmin}} \, {\cal L}(\param \lvert p^{(0)} ) + \lambda_{\R}  \lVert \boldsymbol{\mathrm{D}}_2 \param \rVert_1,
\end{align}
with $\lambda_{\R} > 0$ a \textit{regularization hyperparameter} balancing the regularization term against the data fidelity term. 

The use of the Laplacian operator favors a piecewise linear estimate of $\R$.
In addition, the use of the $L^1$-norm imposes sparsity in the locations where changes in the second derivative actually occur. 
While yielding more realistic estimates than $\widehat \param^{(0)}  $, $\widehat \param^{(1)} $ still lacks significant robustness against the low quality of the data, cf. Fig.~\ref{fig-Rfull} and \cite{Pascal2021}. 

\subsection{Robust to outlier estimation}  
\label{sec-Restb}

A classical approach to address the low quality of the data would
consist in first performing data preprocessing or denoising step
followed by, second, the estimation of $\R$.  This, however, requires
to construct a detailed model for data corruption (missing data and
outliers, pseudo-seasonalities notably).  This is a tedious task as
such a model is likely to be specific to each country, and even likely
to vary for one same country with the different stages of the pandemic
(cf. \cite{Pascal2021}).  Instead, it was proposed in
\cite{Pascal2021} to perform both data denoising and reproduction
number estimation within a single step.  The leading thread is to
modify the functional form in \eqref{eq:estRT} to account for outliers
while staying as close as can be from the pandemic model in
\cite{cori2013new}.  The only assumption on data corruption is that it
can be modeled as sparse outliers $O_t$, i.e., isolated rather than
raws of successive irrelevant values, of unknown values that need to
be estimated in addition to $R_t$.  Ideally, the epidemic model in
\cite{cori2013new} should hence be modified to state that,
conditionally to past counts $\Z_{1:t-1} $ and outliers $\O_{1:t-1} $,
the denoised infection counts $Z_t-O_t$ follow a Poisson distribution,
with non stationary parameter $p^{\star}_t \eqdef R_t \times
\sum_{s = 1}^{\tau_\phi} \Phi_s (Z_{t-s} - O_{t-s})$.
This however results in a functional that would not be jointly convex, which impairs fast and robust minimization \cite{Pascal2021}. 
To preserve convexity, it has instead been proposed in \cite{Pascal2021} to weaken the model into: Conditionally to past counts and Outliers $(\Z_{1:t-1}, \O_{1:t-1})$, $Z_t$ follows a Poisson distribution, with non stationary parameter:  $p^{(2)}_t  \eqdef  (R_t \times \sum_{s = 1}^{\tau_\phi} \Phi_s Z_{t-s}) - O_{t}$.
This leads to estimate $\R$ and $\O$ as, 
\begin{align}
\label{eq:penal_KL}
\nonumber (\widehat{\param}^{(2)}, \widehat{\O}^{(2)}) \eqdef \underset{\R, \O}{\mathrm{argmin}}\, & {\cal L}(\R, \O \lvert  p^{(2)}) + \lambda_{\R}\lVert \boldsymbol{\mathrm{D}}_2 \param \rVert_1 \\
+ \iota_{\geq 0}(\param) + \lambda_{\O} \Vert \O \Vert_1, 
\end{align}
with $\lambda_{\R} > 0$ and $\lambda_{\O} > 0$, \textit{regularization
  hyperparameters}, balancing the strengths of the different
constraints one against each other and against the data fidelity term.
The regularization $\Vert \O \Vert_1 $ favors sparsity in outliers,
the $\{0, +\infty\}$-valued indicator function $\iota_{\geq 0}(\R)$
ensures non-negativity in $\widehat{\param}^{(2)}$.  While robust to
outliers \cite{Pascal2021}, the approximation leading to the criterion
in \eqref{eq:penal_KL} is likely to induce a bias in the estimation of
$\param$, leading to the refinement proposed here.

\subsection{Robust to outliers and unbiased estimation}  
\label{sec-Restc}

To remove the bias described above and hence to further improve estimation, it is proposed here, to assume that conditionally to $\Z_{1:t-1} $ and $\O_{1:t-1} $, $\widehat{Z}^{(D)}_t \eqdef Z_t-\widehat{O}^{(2)}_t$ follows a Poisson distribution with parameter 
\begin{align}
\label{eq:penal_p3}
p^{(3)}_t \eqdef  R_t \times \sum_{s = 1}^{\tau_\phi} \Phi_s (Z_{t-s} - \widehat{O}^{(2)}_{t-s}),
\end{align} thus leading to 
\begin{align}
\label{eq:penal_KLb}
 \widehat{\param}^{(3)} \eqdef \underset{\R}{\mathrm{argmin}}\, & {\cal L}(\param \lvert p^{(3)}) \!+ \lambda_{\R}\lVert \boldsymbol{\mathrm{D}}_2 \param \rVert_1.
\end{align}
In other words, starting from the daily new infection counts, $\Z$, minimization in Eq.~\eqref{eq:penal_KL} is first applied to obtain $\widehat{\O}^{(2)}$, minimization in Eq.~\eqref{eq:penal_KLb} is then applied to the denoised counts $\widehat{\Z}^{(D)} = Z - \widehat{O}^{(2)} $ to yield the final estimate $\widehat \param^{(3)}$. 

\subsection{Credibility intervals from the Metropolis Adjusted Proximal-Gradient Algorithm}  
\label{sec-Restd}

The above penalized Maximum Likelihood procedure provides an optimization-based point estimation of $R$, yet without confidence assessment. 
To complement it, an estimation by means of Credibility Intervals is also proposed in this work. 
It adopts a Bayesian, hence stochastic, perspective on Eq.~\eqref{eq:penal_KLb} and assumes that $ \mathsf{R} $ is a random vector, with a posterior density $ \pi(\param) $  written as \cite{artigas2021credibility}: 
\vspace{-4mm}

\begin{align}
 \label{eq:aposteriori}
 \pi(\param) \propto \exp\bigg(\!\!-\!\!\Big( {\cal L}(\param \lvert p^{(3)}) + \lambda_{\R}\lVert \boldsymbol{\mathrm{D}}_2 \R \rVert_1 + \iota_{\geq 0}(\param) \Big)\bigg).
\end{align}
\vspace{-4mm}

\noindent To produce an estimation of $\R$ by means of credibility
intervals, one resorts to Monte Carlo schemes to produce a sequence of
samples $\{ \param^n \}_{n \geq 0}$ to approximate $\pi$.  The most
classical ones are referred to as Metropolis samplers, and combine two
steps: Proposition and Accept/Reject, as sketched in
Algorithm~\ref{algo:basic}.  When $- \ln \pi$ is a smooth convex
function, the proposition step relies on Langevin
dynamics~\cite{parisi:1981,roberts:tweedie:1996}, whose key idea is to drive the
proposition with the gradient $\nabla$ of $ \ln \pi$, as
\begin{equation} \label{eq:Langevin:dynamic}
\mu(\param^n) \eqdef  \param^n + \pas \Gamma \nabla \ln \pi(\param^n),
\end{equation}
and to perturb it with an additive correlated Gaussian noise $\sqrt{2
  \pas} \Gamma \epsilon_{n+1}$, where $\epsilon_{n+1} \sim
\mathcal{N}(0_T, \I_T)$; $\pas$ is a positive step size.
The accept/reject step relies on a Metropolis mechanism,
cf. Eq.~\eqref{eq:ARratio}. We set $q(\param, \param') \eqdef
\mathcal{N}\left(\mu(\param), 2 \pas \Gamma \Gamma^\top \right)
        [\param'] \eqsp$.
  \vskip-3mm
\begin{algorithm}
  \KwInput{$N_{\mathrm{max}} \in \nset_\star$, $\param^{0}$, $\gamma>0$, $\Gamma$} 
  \KwOutput{$\{\R^n\}_{n=0}^{N_{\mathrm{max}}} $ } 
  \For{$n=0, \ldots, N_{\mathrm{max}}-1$}{
   Step1: Draw 
     $\param^{n+1/2} \sim \mu(\param^{n})  + \sqrt{2 \pas} \, \Gamma \epsilon_{n+1}$ \;
   Step2:  $\param^{n+1} = \param^{n+1/2}$ with probability
    \begin{equation} \label{eq:ARratio}
1 \wedge \frac{\pi(\param^{n+1/2})}{\pi(\param^n)}
\frac{q(\param^{n+1/2}, \param^n)}{q(\param^n, \param^{n+1/2})}
    \end{equation}
    and $\param^{n+1} = \param^n$ otherwise.  \label{line:AR}
  }
  \caption{Metropolis Adjusted Proximal-Gradient Algorithm \label{algo:basic}} 
  \vskip-1mm
\end{algorithm}

In the case of Eq.~\eqref{eq:aposteriori}, because of the
regularization terms, $- \ln \pi$ is a convex but non smooth function.
To preserve the key intuition of the Langevin scheme
\eqref{eq:Langevin:dynamic}, the gradient step is replaced with a
proximal-gradient step, a suited extension to 
nonsmooth
functions.  Several developments were conducted in that line
\cite{schreck:etal:2016,durmus:etal:2018,durmus:etal:2019,salim:richtarik:2020}.
Recently, we proposed in \cite{artigas2021credibility}, a sampling
scheme well suited to the structure and properties of $\pi$ in
Eq.~\eqref{eq:aposteriori}.  It amounts to write the drift $\mu$ in
the Gaussian proposition:
\begin{equation} \label{eq:}
\mu(\param^n) \eqdef  \overline{\mathsf{D}}_o^{-1} \Prox_{\pas \lambda_{\R} \| \cdot \|_1}\left(\overline{\mathsf{D}}_o \param^n - \pas \overline{\mathsf{D}}_o^{-\top}  \nabla {\cal L}(\param^n,p^{(3)}) \right)
\end{equation}
where $\overline{\mathsf{D}}_o$ is a $ T \times T$ invertible matrix
obtained by orthogonal complementation of the $ (T -2) \times T$
Toeplitz matrix associated with the Laplacian
$\mathsf{D}_2$ and $\Prox_{\pas \lambda_{\R} \| \cdot
  \|_1}$ is a soft thresholding operation.  Full technical details are
provided in \cite{artigas2021credibility,abry:etal:2022}.

\section{Covid19 Data}
\label{sec-data}

The \emph{Johns Hopkins University} has developed and maintains a remarkable Covid19 data repository, \href{https://coronavirus.jhu.edu/}{https://coronavirus.jhu.edu/}, impressively started with the outbreak of the pandemics. 
It notably collects on a daily basis new infection and death counts, as produced by the National Health Authorities of around 200 countries or autonomous territories, and makes them publicly available in a consistent setting.
This constitutes an exceptional source of data for the Covid19 pandemic monitoring and intensity assessment, as data are made available in (quasi-)real time and within the pandemics. 
Because the present work focuses on the reproduction number, daily new infection counts only are used here. 
Figs.~\ref{fig-Rfull} and ~\ref{fig-Rshort} (top plots) illustrate the time evolution of such counts for several countries. 

\section{Reproduction number estimation} 
\label{sec-Res}

\begin{figure}[t]
\centerline{
\includegraphics[width=0.5\linewidth]{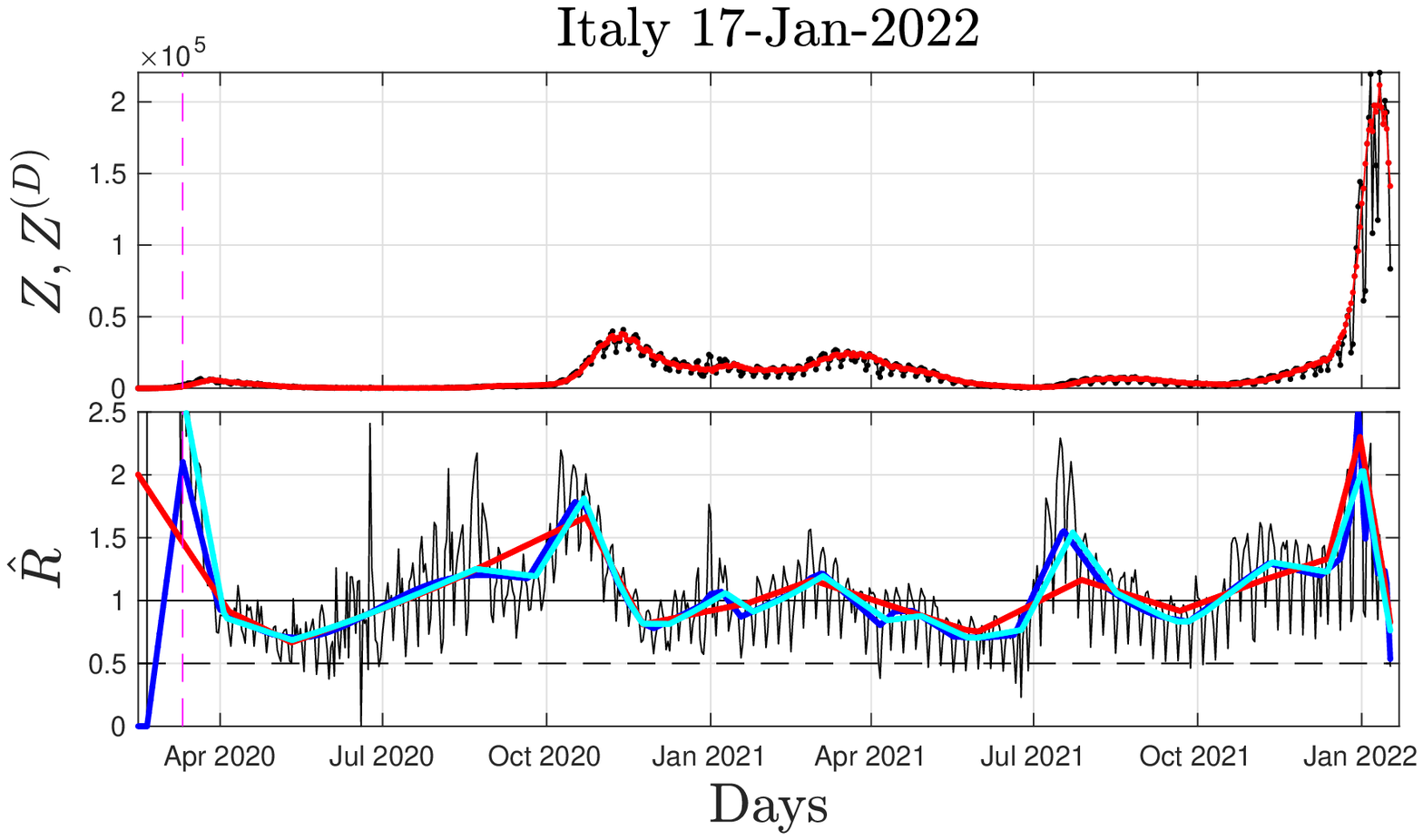}
\includegraphics[width=0.5\linewidth]{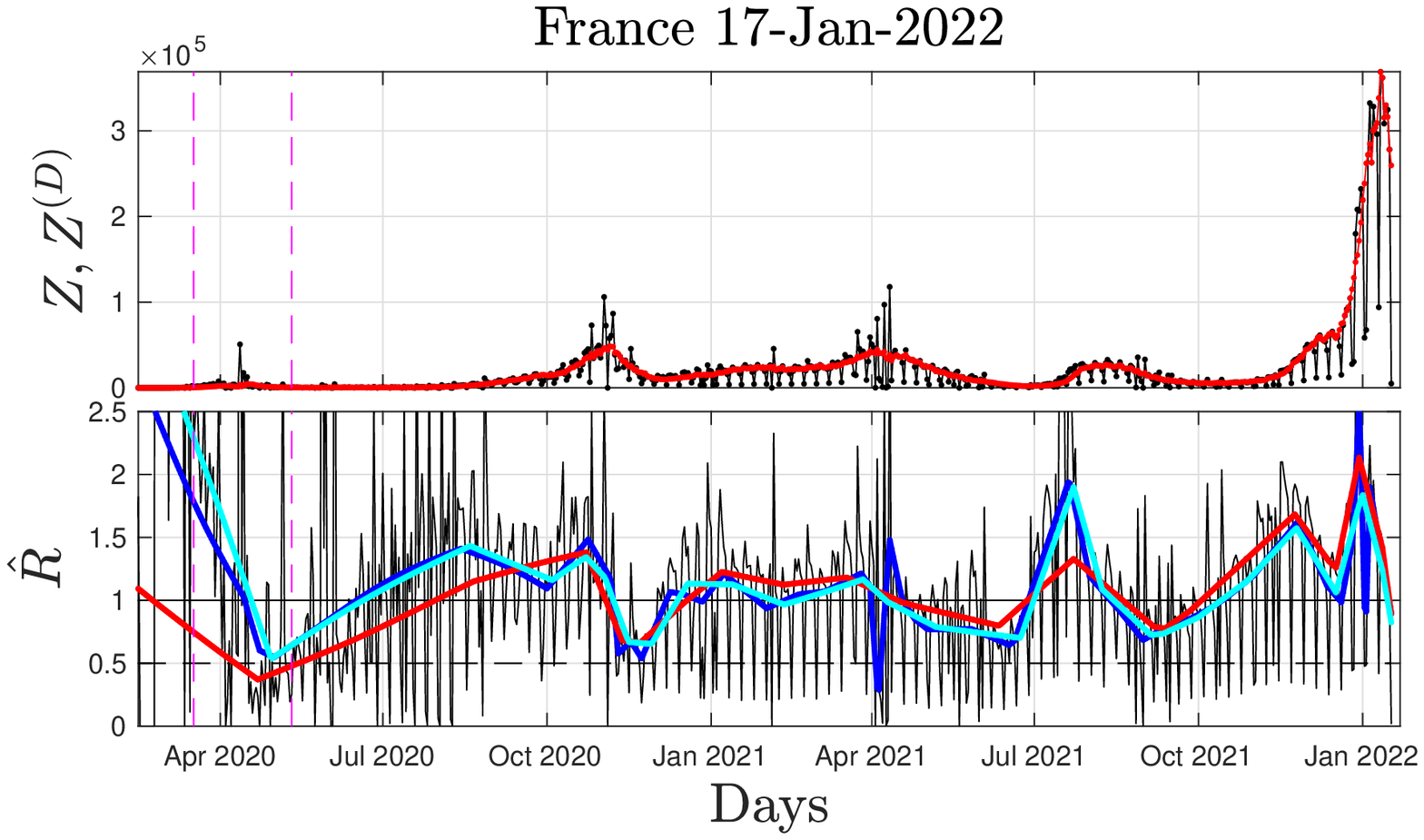}
}
\centerline{
\includegraphics[width=0.5\linewidth]{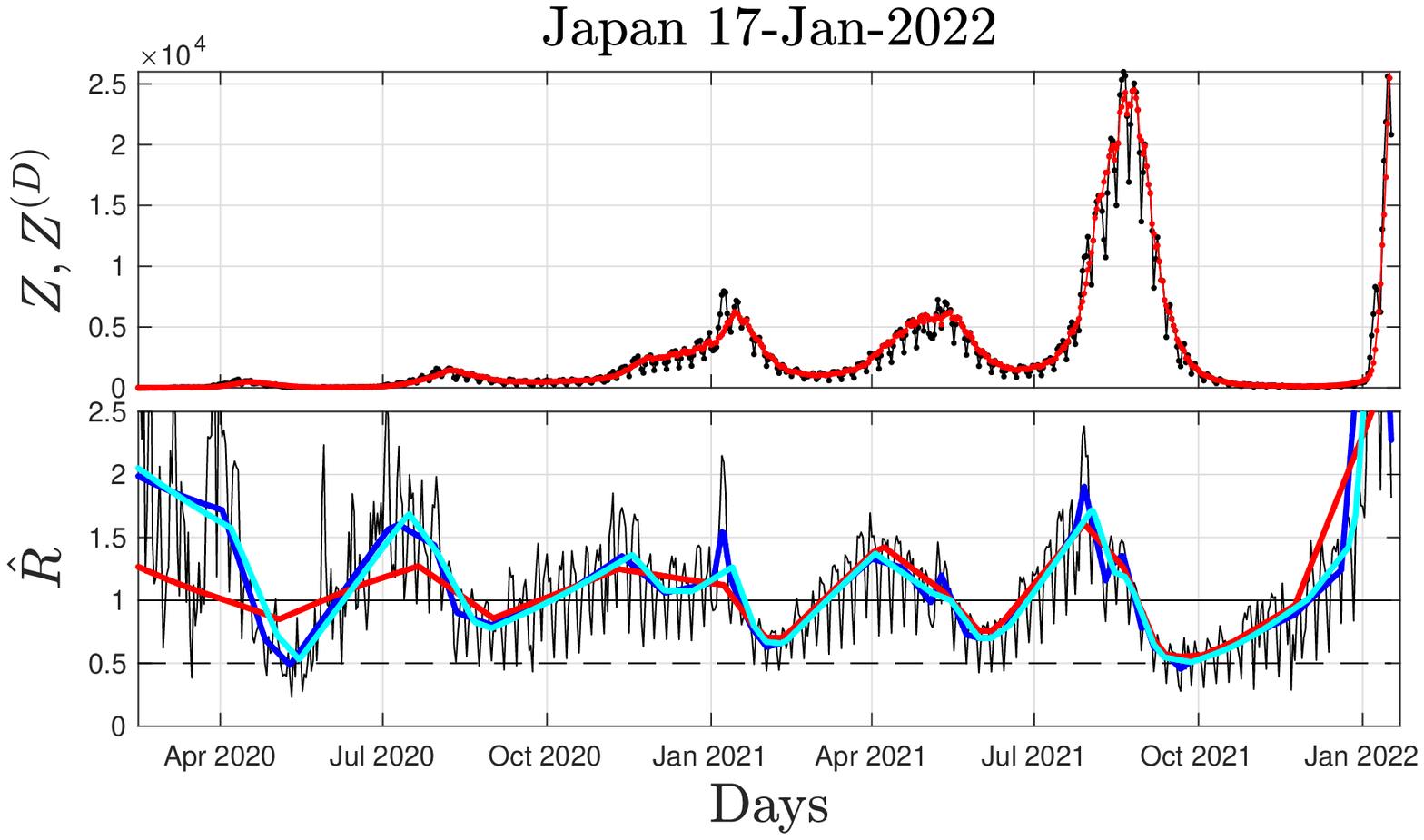}
\includegraphics[width=0.5\linewidth]{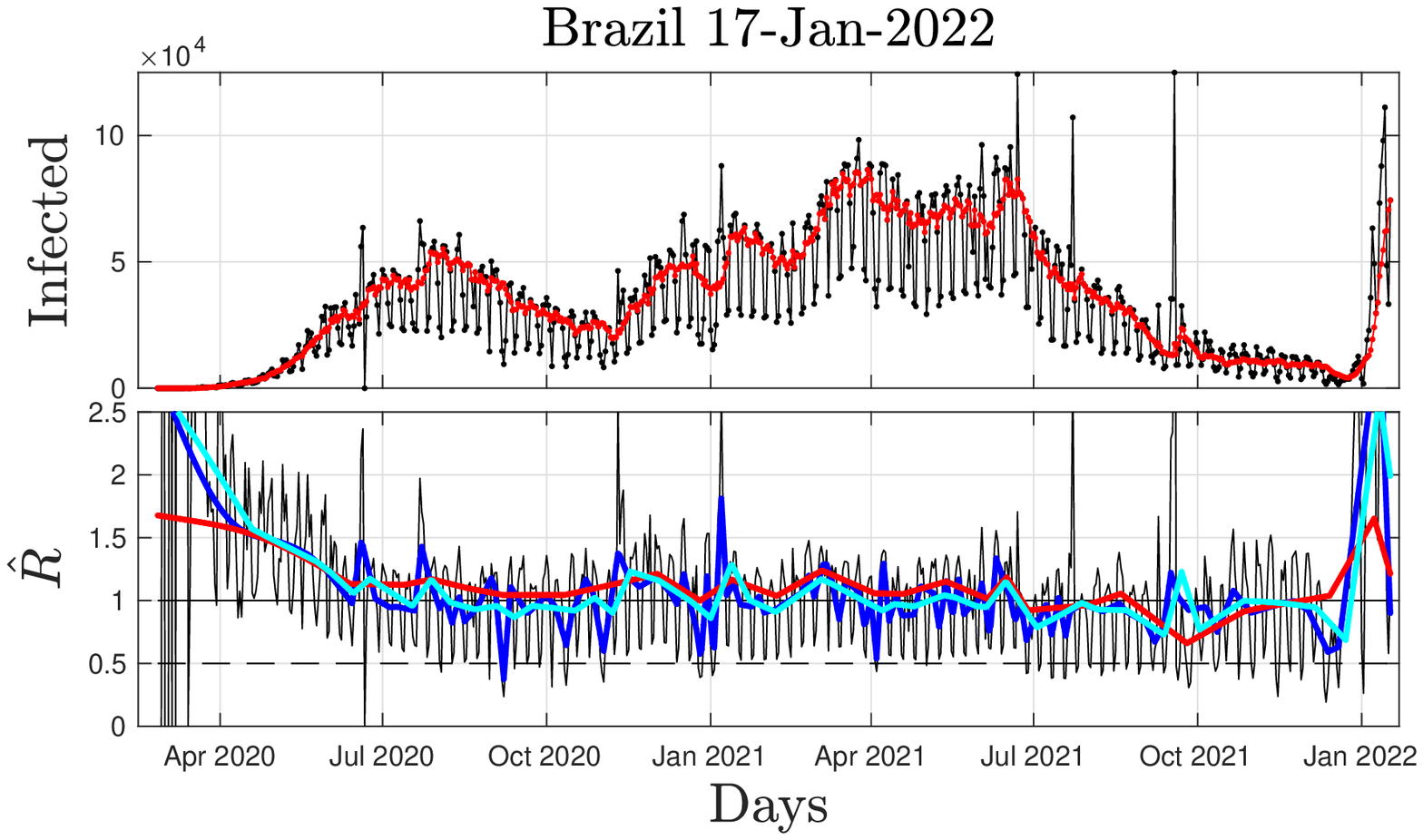}
}
\vspace{-4mm}
\caption{\label{fig-Rfull}{\bf Reproduction number estimations for the entire pandemic period, for four different countries.} Top: Raw ($\Z$, black) and denoised  ($\widehat{\Z}^{(D)}$, red) daily new infection counts. Bottom: estimates for $\param$, $\widehat \param^{(0)} (black), \widehat \param^{(1)} (blue),  \widehat \param^{(2)} (red), \widehat \param^{(3)} (cyan)$.} 
\vspace{-0.4cm}
\end{figure}

Outcomes of the proposed estimation procedures are illustrated for a
few countries for space reasons.  Yet, procedures are operational for
any country.  Daily updated estimations are automatically made
available at 
\href{https://perso.ens-lyon.fr/patrice.abry/}{perso.ens-lyon.fr/patrice.abry/} and \href{https://www.math.univ-toulouse.fr/~gfort/}{www.math.univ-toulouse.fr/gfort/}.
Following \cite{artigas2021credibility,Pascal2021}, the hyperparameter
are set to $\lambda_0 = 0.05$, $\lambda_{\R} = 3.5 \times $ std$(\Z)$/4
with std$(\Z)$ the standard deviation of $\Z$. \\
\noindent {\bf Inverse problem estimation.} Fig.~\ref{fig-Rfull} reports (top row) raw ($\Z$) and denoised ($\widehat{\Z}^{(D)}$) daily counts of new infections for the full period of the pandemic.
Fig.~\ref{fig-Rfull} compares (bottom row) the different estimates proposed here: $\widehat \param^{(0)}, \widehat \param^{(1)},  \widehat \param^{(2)}, \widehat \param^{(3)}$, leading, for all countries, to the following conclusions: 
The crude estimator $\widehat \param^{(0)}$ yields estimates that are far too irregular in time to be useful by epidemiologists. 
The time regularized estimator $\widehat \param^{(1)}$ consists of piecewise linear estimates of $\param$ and thus provides far more regular and hence realistic assessments of the pandemic intensity evolution. 
Yet, Fig.~\ref{fig-Rfull} also shows that $\widehat \param^{(1)}$ lacks robustness against outliers and missing counts in the raw $Z$. 
Further,  \eqref{eq:penal_KL} jointly provides estimates of outliers $\widehat \O^{(2)} $ and piecewise linear estimates $\widehat \param^{(2)} $ of $\param$ that are robust to irrelevant counts in $\Z$, and thus of far greater interest to epidemiologists. 
Yet, a careful examination of the approximation made in the outlier modeling to maintain the convexity of the functional suggests a possible overestimation in  $\widehat \param^{(2)}$.
Finally, $\widehat \param^{(3)}$, obtained from the denoised counts $\widehat{\Z}^{(D)}$, provides the most relevant and useful estimation of $\param$, with smooth (piecewise linear) and  accurate estimations of $\param$, permitting notably the detection of the occurrences of changes between pandemic growth and regression phases.

\begin{figure}[t]
\centerline{
\includegraphics[width=0.5\linewidth]{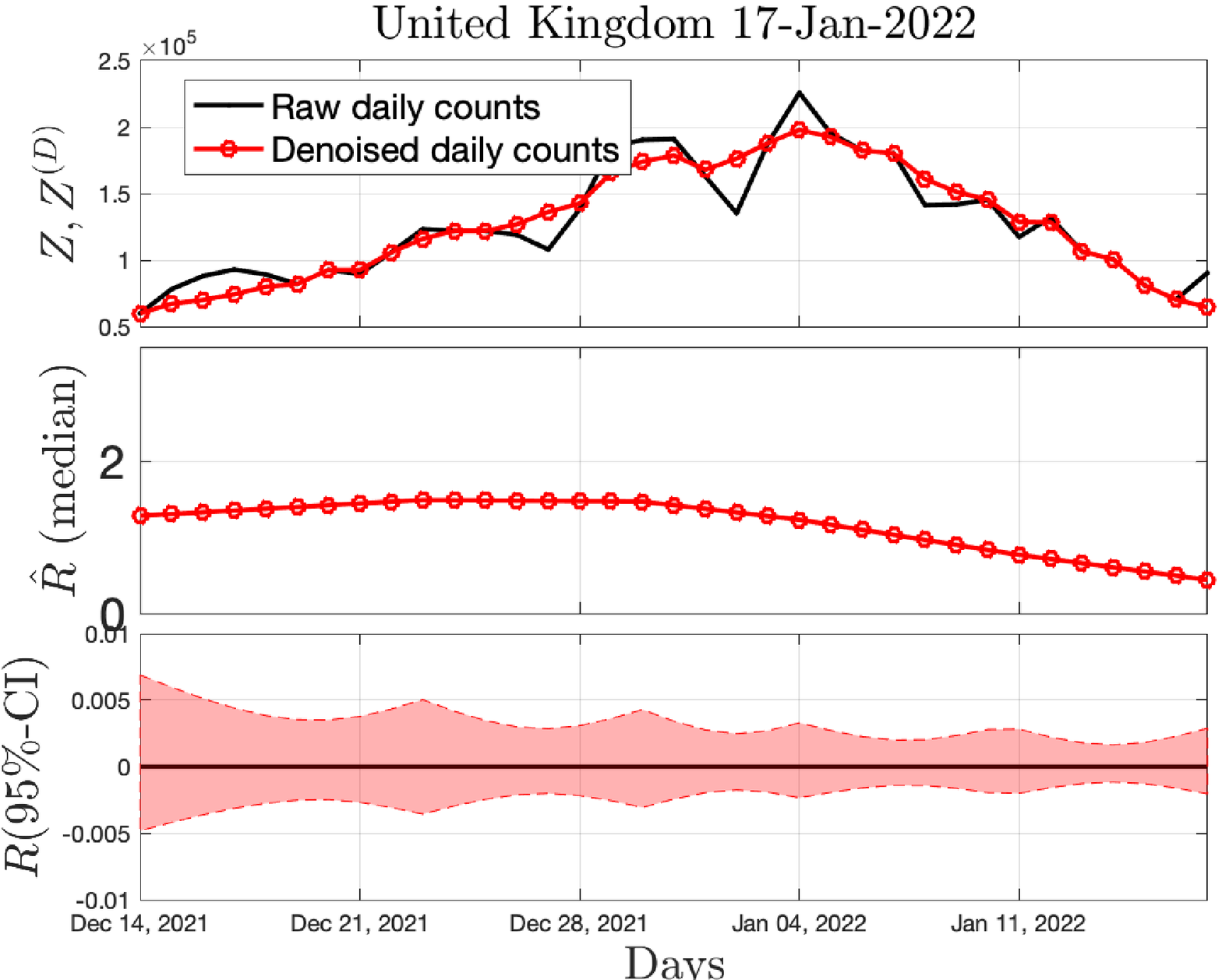}
\includegraphics[width=0.5\linewidth]{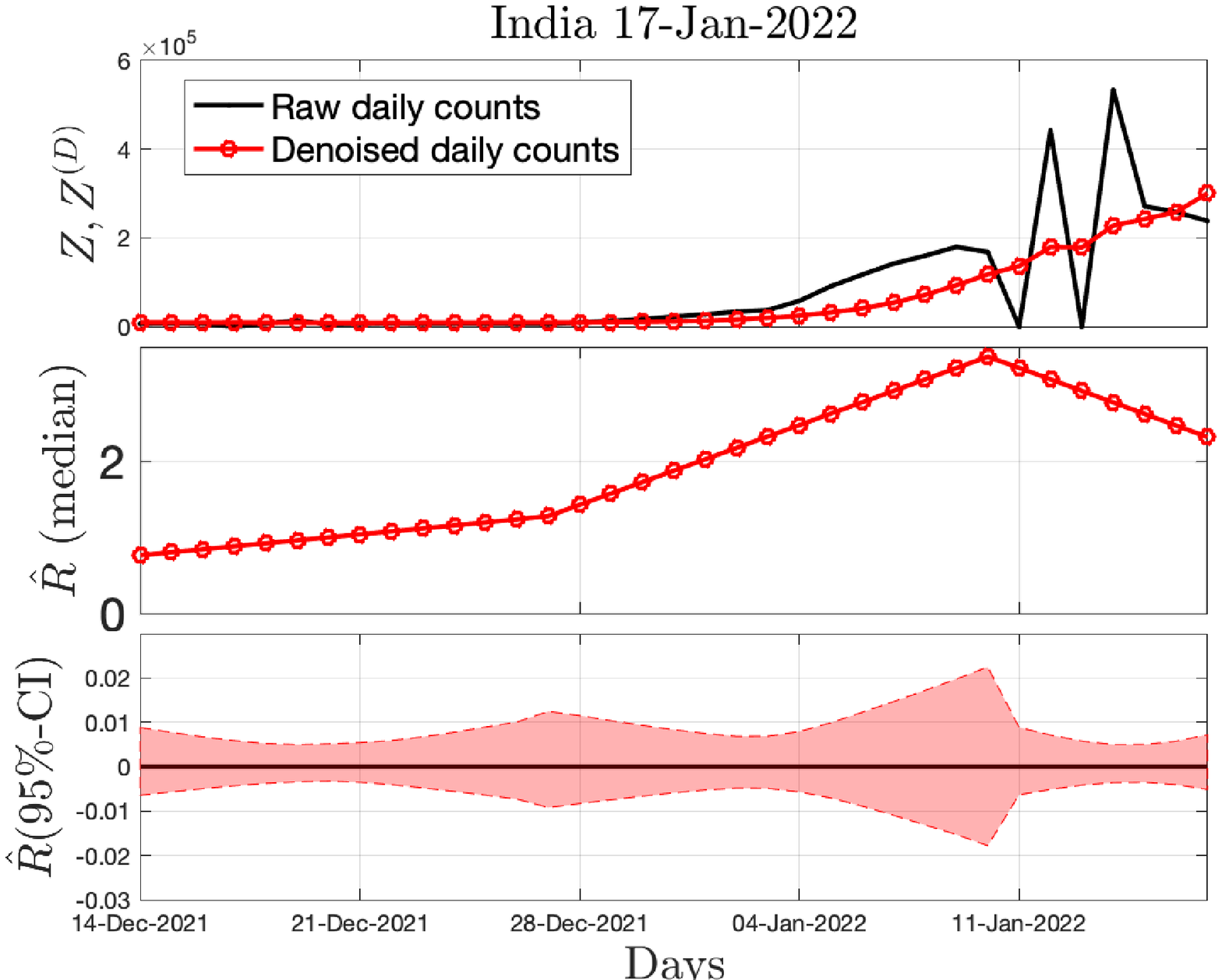}
}
\centerline{
\includegraphics[width=0.5\linewidth]{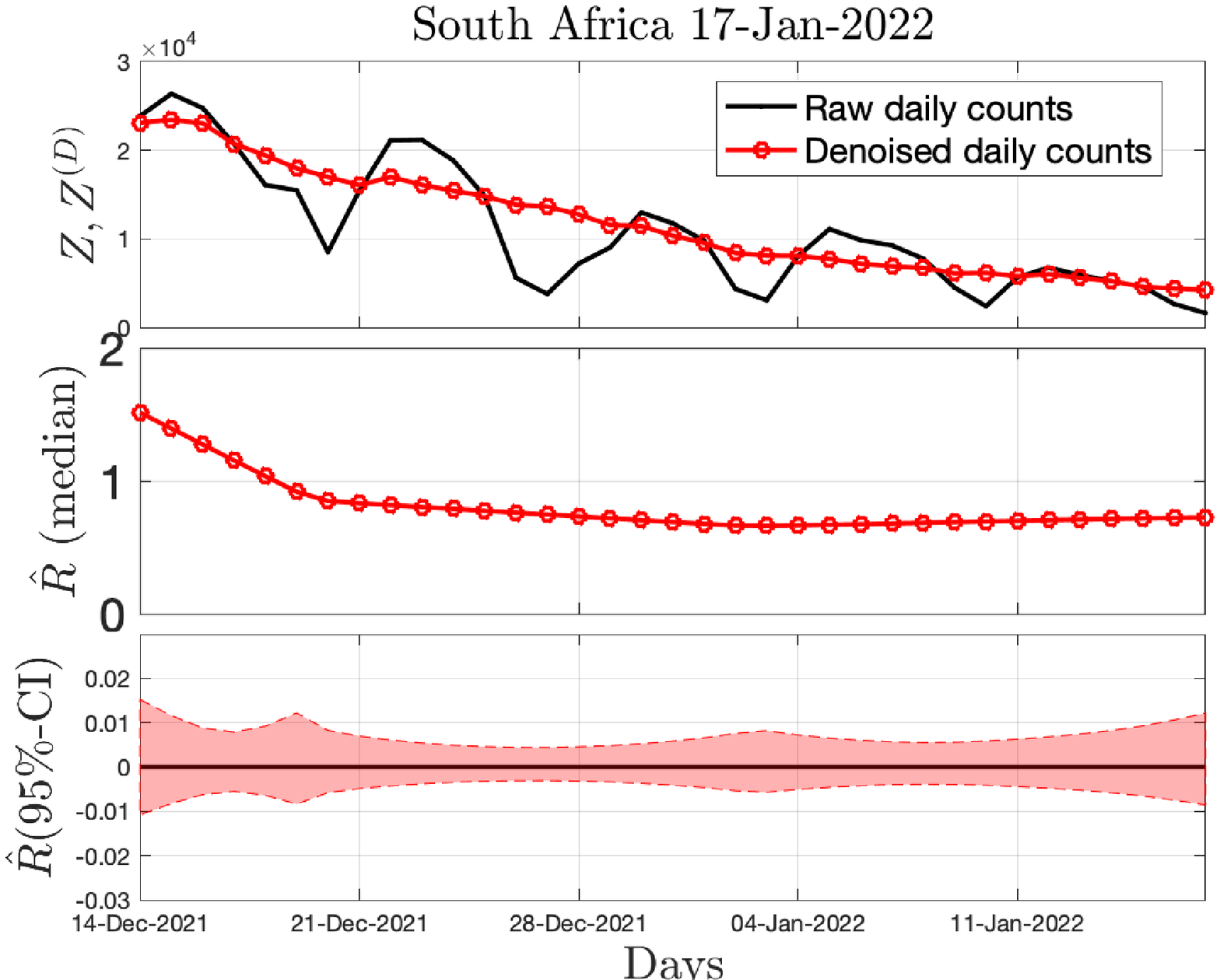}
\includegraphics[width=0.5\linewidth]{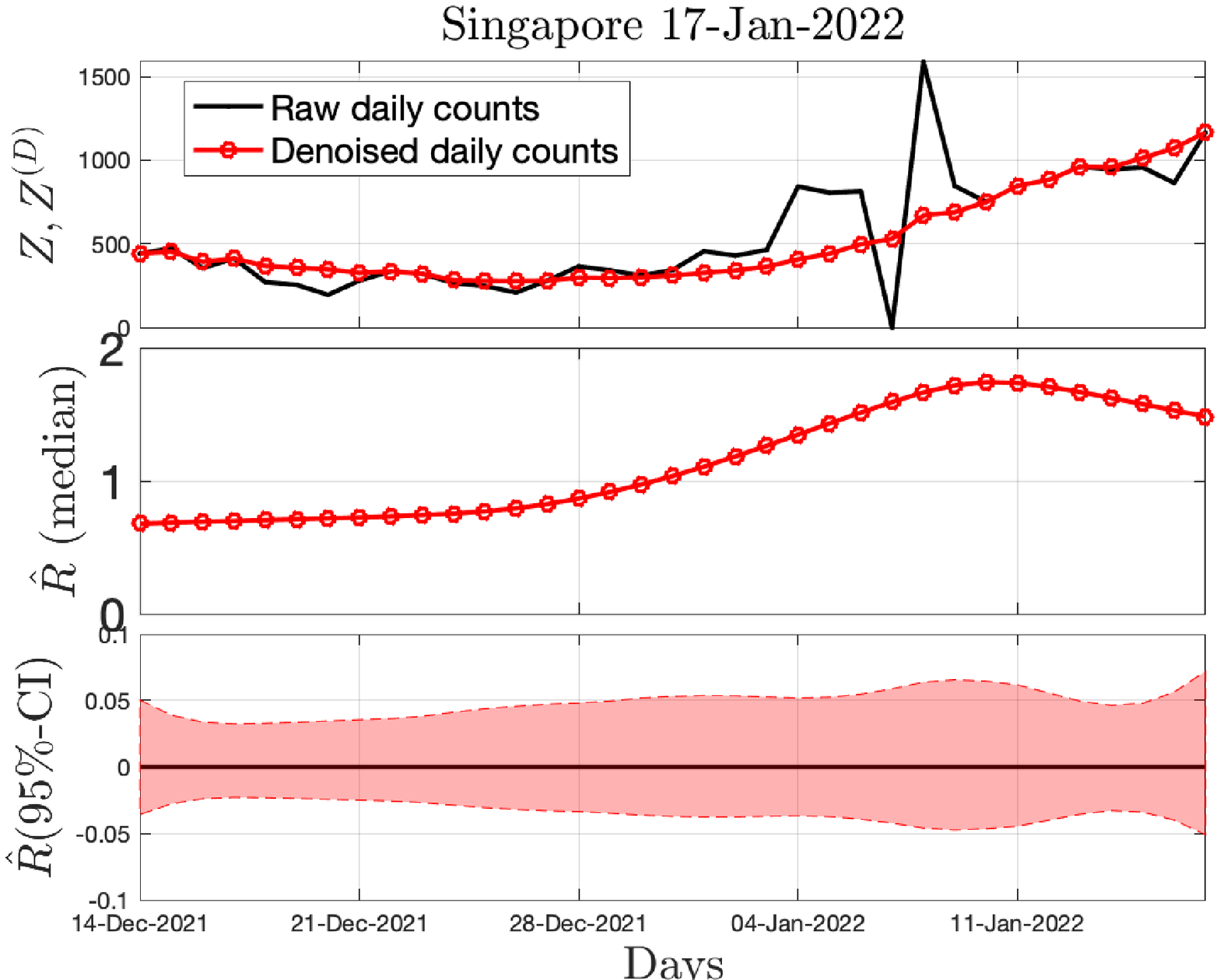}
}
\vspace{-4mm}
\caption{\label{fig-Rshort}{\bf Credibility interval estimation for
    the reproduction number estimations for the 35 last days and four
    different countries.}  Top: Raw ($\Z$, black) and denoised
  ($\widehat{\Z}^{(D)}$, red) daily new infection counts.  Middle: a
  posteriori median ($50\%$-quantile) estimate for $\param$.  Bottom:
  $95 \%$-credibility interval estimate for $\param$, reported as the
  plots of the $97.5\%$ and $2.5\%$-quantiles, after subtraction of
  the $50\%$-quantile.}
\vspace{-0.6cm}
\end{figure}

\noindent {\bf Credibility interval estimation.} Fig.~\ref{fig-Rshort}
focuses on the most recent five weeks of the pandemic.  It reports raw
($\Z$) and denoised ($\widehat{\Z}^{(D)}$) new infection daily counts
for other countries (top raw).  It shows (middle row) the estimated
median ($50\%$-quantile) of the posterior distribution sampled by the
strategy described in Section~\ref{sec-Restd} and the corresponding
centered $95\%$ credibility intervals, obtained from the $2.5\%$ and
$97.5\%$-quantiles and after subtraction of the $50\%$-quantile
(bottom row), leading for all countries, to the following conclusions:
The credibility intervals are extremely narrow (around a few $\%$)
around the median, and relatively homogeneous along time, yet with
mild increase around the piecewise linearity change points.

\section{Conclusions and perpectives} 

These results show that both the inverse problem formulations and the
Metropolis Adjusted Proximal-Gradient sampler proposed here yields
extremely realistic estimates for the time evolution of $\param$, that
are hence actually usable by epidemiologists.  Notably, these
estimation tools have a double potential value: Retrospectively, they
permit to quantify the impacts of given sanitary measures on the
pandemic evolution~; Prospectively, the piecewise linear nature of the
estimation of $\param$ permits the short term forecast (the
\emph{nowcast}) of the evolution of the pandemic intensity.  Further,
sampling strategies for Credibility Interval joint estimation for both
the reproduction number $\param$ and the Outliers $\O$ are being
devised and compared, with several formulations of convex nonsmooth
compliant Proposition steps (cf. \cite{abry:etal:2022}).

Finally, these estimation tools are being made publicly available in a document toolbox, as a contribution to open science and dedication of science to major societal stakes.

\bibliographystyle{IEEEbib}

\end{document}